\documentclass{aastex62}

\graphicspath{{./}{}}

\received{January 1, 2018}
\revised{January 7, 2018}
\accepted{\today}
\submitjournal{ApJ}

\shorttitle{Rubin's Galaxy Nucleus}
\shortauthors{Holwerda et al.}

\begin{document}

\title{Predicting the spectrum of UGC 2885, Rubin's Galaxy with machine learning.}

\correspondingauthor{Benne W. Holwerda}
\email{benne.holwerda@louisville.edu}

\author[0000-0002-0786-7307]{Benne W. Holwerda}
\affil{Physics \& Astronomy Department, University of Louisville, 40292 KY, Louisville, USA}

\author[0000-0002-5077-881X]{John F. Wu}
\affiliation{Space Telescope Science Institute, 3700 San Martin Drive, 21219 MD Baltimore, USA}

\author[0000-0002-6131-9539]{William C. Keel}
\affil{Department of Physics and Astronomy, University of Alabama Box 870324, Tuscaloosa, AL 35487-0324, USA}

\author[0000-0002-5830-9233]{Jason Young}
\affil{Mount Holyoke College, 50 College Street, South Hadley, MA 01075, USA}

\author{Ren Mullins}
\affil{Physics \& Astronomy Department, University of Louisville, 40292 KY, Louisville, USA}

\author[0000-0003-3339-0546]{Joannah Hinz}
\affil{Steward Observatory, University of Arizona, 933 North Cherry Avenue, Tucson, AZ 85721- 0065, USA}
\affil{MMT Observatory, P.O. Box 210065, Tucson, AZ 85721-0065, USA}

\author[0000-0002-5956-851X]{K.E. Saavik Ford}
\affil{Dept. of Science, Borough of Manhattan Community College, City University of New York, New York, NY 10007 USA}
\affil{Dept. of Astrophysics, American Museum of Natural History, New York, NY 10024 USA}
\affil{Physics Program, CUNY Graduate Center, City University of New York, New York, NY 10016 USA}
\affil{Center for Computational Astrophysics, Flatiron Institute, New York, NY, 10010, USA}

\author[0000-0003-2767-0090]{Pauline Barmby}
\affil{Department of Physics \& Astronomy, Institute for Earth \& Space Exploration, University of Western Ontario, 1151 Richmond Street, London, Ontario, Canada}

\author[0000-0003-0085-4623]{Rupali Chandar}
\affil{Department of Physics \& Astronomy, The University of Toledo, Toledo, OH 43606, USA}

\author[0000-0001-6380-010X]{Jeremy Bailin}
\affil{Department of Physics and Astronomy, University of Alabama Box 870324, Tuscaloosa, AL 35487-0324, USA}

\author[0000-0003-4797-7030]{Josh Peek}
\affiliation{Space Telescope Science Institute, 3700 San Martin Drive, 21219 MD Baltimore, USA}
\affiliation{Department of Physics \& Astronomy, Johns Hopkins University, Baltimore, MD 21218, USA}

\author[0000-0002-9427-5448]{Tim Pickering}
\affil{Steward Observatory, University of Arizona, 933 North Cherry Avenue, Tucson, AZ 85721- 0065, USA}
\affil{MMT Observatory, P.O. Box 210065, Tucson, AZ 85721-0065, USA}

\author[0000-0002-5666-7782]{Torsten B\"oker}
\affiliation{European Space Agency, c/o STScI, 3700 San Martin Drive, 21219 MD Baltimore, USA}



\begin{abstract}
\cite{Wu20a} predict SDSS-quality spectra based on Pan-STARRS broad-band \textit{grizy} images using machine learning (ML). In this letter, we test their prediction for a unique object, UGC 2885 (``Rubin's galaxy''), the largest and most massive, isolated disk galaxy in the local Universe ($D<100$ Mpc).
After obtaining the ML predicted spectrum, we compare it to all existing spectroscopic information that is comparable to an SDSS spectrum of the central region: two archival spectra, one extracted from the VIRUS-P observations of this galaxy, and a new, targeted MMT/Binospec observation.
Agreement is qualitatively good, though the ML prediction prefers line ratios slightly more towards those of an active galactic nucleus (AGN), compared to archival and VIRUS-P observed values. The MMT/Binospec nuclear spectrum unequivocally shows strong emission lines except H$\beta$, the ratios of which are consistent with AGN activity. The ML approach to galaxy spectra may be a viable way to identify AGN supplementing NIR colors.
%
How such a massive disk galaxy ($M^* = 10^{11}$ M$_\odot$), which uncharacteristically shows no sign of interaction or mergers, manages to fuel its central AGN remains to be investigated.

\end{abstract}



\section{Introduction} \label{sec:intro}

In recent years, astronomical machine learning has transformed from a niche subject to a well-developed suite of accessible and interpretable algorithms. Machine learning algorithms are used for catalogs, time series, imaging, and unstructured data across a wide array of astronomy subdisciplines (e.g., exoplanets, \citealt{Shallue18}; radio interferometry, \citealt{Vafaei-Sadr20}; transient photometry, \citealt{Villar20}; and studies of the interstellar medium, \citealt{Xu20}). For extragalactic imaging data, machine learning is regularly relied upon for classifying galaxy morphology \citep[e.g.,][]{Dieleman15,Beck18}, quantifying observed properties \citep[e.g.;][]{Smith20a}, and predicting spectroscopic properties \citep[e.g.,][]{Pasquet19,Wu19b}. These advances have enabled machine learning to take on an outsized role in mitigating the observational disparity between enormous survey imaging data sets and limited spectroscopic follow-up campaigns \citep[e.g,][]{Wu20a}. While machine learning results are robust for estimating the properties of \textit{typical} galaxies, it is less clear if predictions are reliable for rare or unusual systems.


Spectroscopy surveys lag behind imaging ones in coverage and depth simply because of the longer integration time per object \citep[see][for a discussion on ongoing and near-future surveys]{Davies18,Driver19}. Predicting spectra or spectral features based on imaging is therefore a worthwhile endeavour for both scientific and planning purposes.

UGC 2885 (``Rubin's Galaxy") is the largest and most massive spiral galaxy ($R_p = 38.1$~kpc, $M_* = 2.2\times 10^{11}$ M$_\odot$) in the local Universe \citep{Rubin80b}. It shows very regular rotation with a maximum rotation velocity of 298 km~s$^{-1}$ \citep[$ M_{tot} = 1.50\times10^{12} $ M$_\odot$][]{Rubin80a,Roelfsema85,Wiegert14,McGaugh15a}, rising from the center, indicating a substantial and extended dark matter halo \citep{Marr15a}. The galaxy has  star-formation throughout its disk \citep{Hunter13}, four spiral arms, and a specific incidence of globular clusters that hints at an uneventful accretion history (Holwerda et al in prep.). 
It lies just outside the Sloan Digital Sky Survey's footprint and has no SDSS or DES spectra. 

UGC 2885 defies easy classification: to the first order it is an Sc galaxy but it is at the outermost edge of the mass and size envelope for its class. 
The initial identification of UGC 2885 as the largest isolated spiral galaxy by \cite{Rubin80b} was confirmed by \cite{Romanishin83} for the whole Uppsala Galaxy Catalog. We measure the Petrosian radius of Rubin's galaxy at 38.1 kpc (97\") in sdss-r. 
\cite{Ogle16,Ogle19} introduce ``Super spirals,'' a class of massive disk galaxies at the utmost top of the stellar mass-star formation rate ($M_*$-SFR) relation. \cite{Ogle15} note that ``Super spirals" are almost never in an isolated environment and often show signs of a recent interaction, e.g. perturbed morphology, dual nuclei, or other signs of significant merger activity.  
Rubin's galaxy does not meet these qualifications as it has a lower SFR and does not show signs of recent interaction, which appear to be typical for this class of galaxies. \cite{Saburova20a} discusses giant low surface brightness (LSB) disk galaxies. Once again, UGC 2885 does not meet those specifications either: its star-formation is too high and concentrated in a more typical brightness disk. These giant LSB galaxies also often seem to be visibly disturbed from recent accretion or merger activity. Yet Rubin's galaxy is remarkably unperturbed. 

A $\sim10^{11} M_\odot$ stellar mass galaxy typically has an active galactic nucleus and our question for this letter is: is Rubin's galaxy exceptional here as well? \cite{Keel83a} remains the only study to attempt to characterize the nucleus of this galaxy and does not reach a firm conclusion. Does UGC 2885 show signs of an active nucleus? One would expect its spectrum to lie in the categories of LINER or star-formation/AGN mix. 
Can this be predicted from broad-band photometry using machine learning? How good is the ML prediction of this galaxy's spectrum? As a local extreme of size and mass, the lack of a SDSS spectrum offered the opportunity for a direct test of this prediction.

The purpose of this paper is to report a ``closed envelope" test of the ML predicted spectrum for the center of Rubin's galaxy from \cite{Wu20a}  and compare it to spectra obtained from private archives and recent observations: the prediction was made first and then compared to spectra of comparable quality.

\section{Machine Learning Prediction of Rubin's Galaxy's Spectrum} \label{sec:ML}

\begin{figure}
    \centering
    \includegraphics[width=\textwidth]{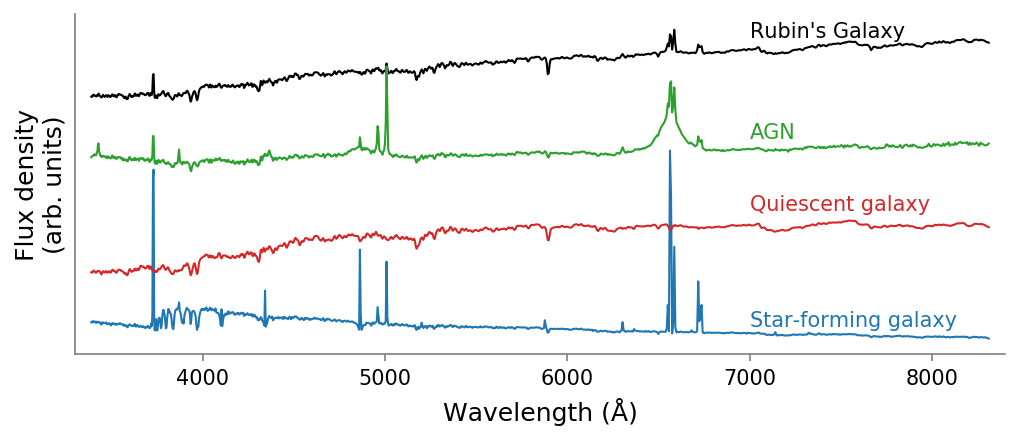}
    \caption{The prediction by the ML method of \cite{Wu20a} for Rubin's Galaxy (top) and reference SDSS spectra for an example of an AGN, a quiescent galaxy, and a star-forming galaxy, all shifted to the rest frame. The spectral features of Rubin's Galaxy fall between those of a quiescent and an AGN-dominated galaxy.}
    \label{f:ML-spectrum}
\end{figure}

We generate the predicted spectrum for Rubin's Galaxy using the method described in \cite{Wu20a}. Briefly, the method uses a convolutional neural network (CNN), trained on $56\arcsec \times 56\arcsec$ $grizy$ image cutouts of galaxies from the Pan-STARRS $3\pi$ Steradian Survey DR2, and predicts the galaxies' optical spectra (3$\arcsec$ aperture, SDSS resolution and s/n). Galaxy spectra are normalized, shifted to a common reference frame, and rebinned in 1000 logarithmically spaced elements. A variational autoencoder (VAE) represents each spectrum using six latent variables, which can be robustly ``decoded'' back into SDSS spectra \citep{Portillo20}. \cite{Wu20a} train a CNN to predict galaxies' SDSS spectra solely using image cutouts.  Figure \ref{f:ML-spectrum} shows the results of this ML approach applied to the Pan-STARRS image of Rubin's galaxy. In the comparisons this spectrum is shifted to the galaxy redshift of $v = 5802$~km~s$^{-1}$ \citep{Canzian93}.


\section{Data} \label{sec:data}

\begin{figure}
    \centering
    \includegraphics[width=\textwidth]{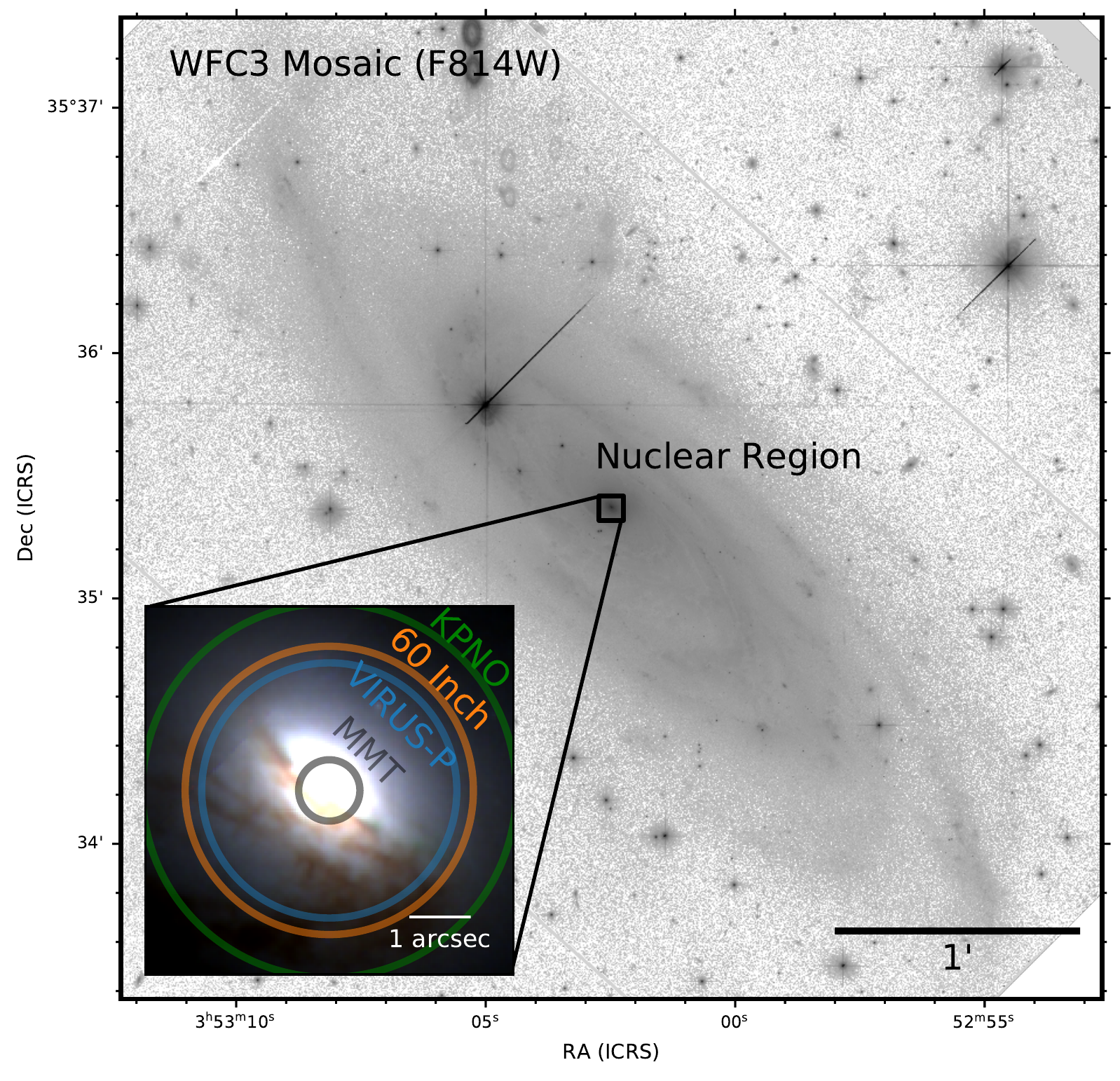}
    \caption{The nucleus of Rubin's galaxy from the larger HST mosaic. A clear dustlane partially obscures the central stellar cluster. The inset shows the apertures of the spectroscopic observations are marked using the same color scheme as in Figure \ref{f:spectra:comparison}: Mount Lemon 60inch (orange), 2.1 KPNO (green), VIRUS-P (blue) and MMT (white for contrast). }
    \label{f:Hubble:nucleus}
\end{figure}

To verify the ML prediction, we obtained two archival spectra, extracted one from the VIRUS-P IFU observation, and obtained new MMT/Binospec observations. The apertures used in each observations is shown in Figure \ref{f:Hubble:nucleus}. 

The first comparison spectrum is originally from \cite{Keel83a,Keel83b}, who conducted an aperture spectroscopic survey of nearby galaxies to search for AGN activity. \cite{Keel83a} reported line strengths for H$\alpha$, [{N}{ii}] and [{S}{ii}]. Their verdict on nuclear activity was inconclusive. 

\begin{figure*}
    \centering
    \includegraphics[width=\textwidth]{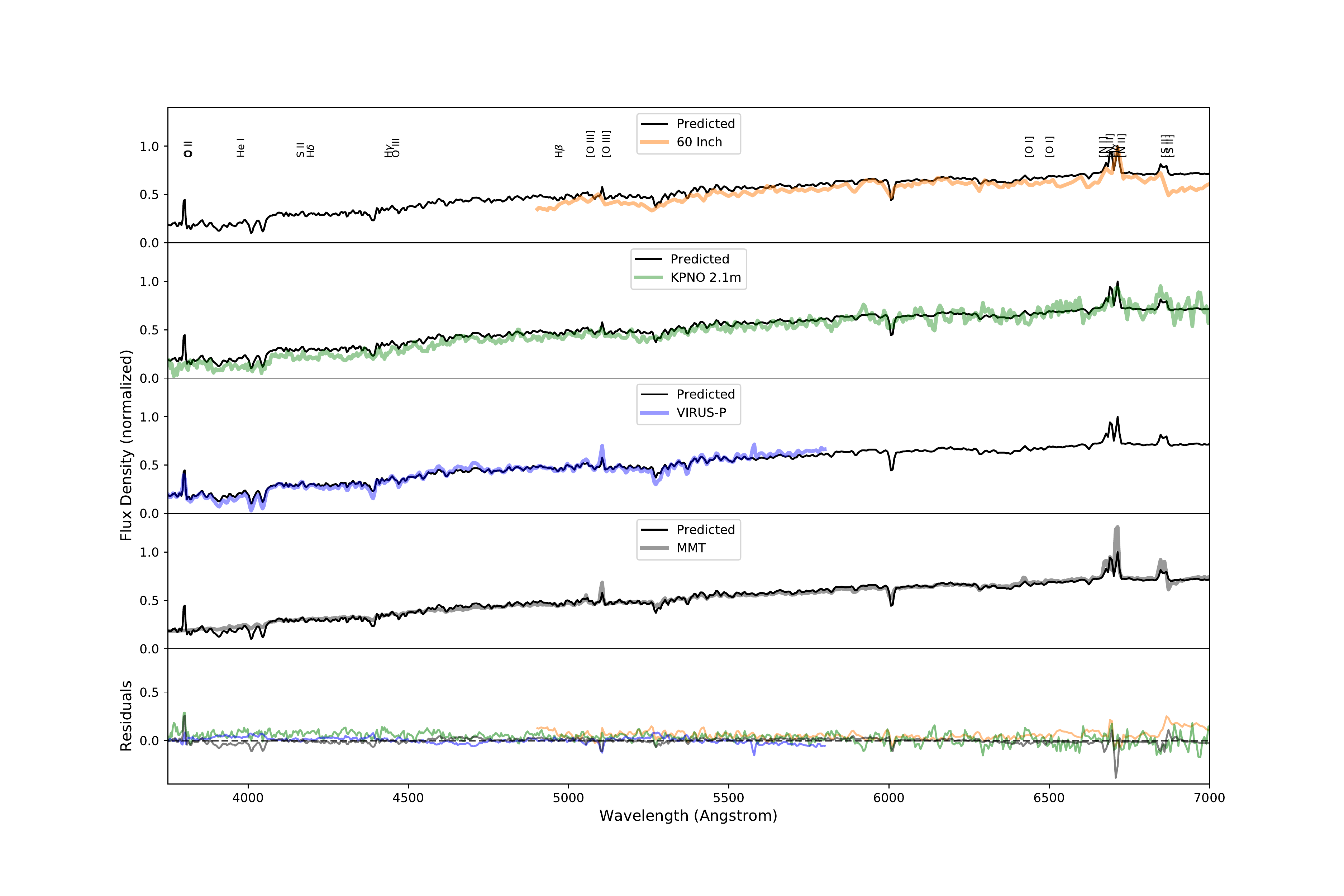}
    \caption{The ML-predicted spectrum (black line in top four panels) and the observed spectra, from 
    \cite{Keel83a} using the Mount Lemon 60 inch (top panel, orange), 
    the KPNO 2.1~m telescope (second panel, green), 
    from the VIRUS-P IFU instrument (third panel, blue), 
    and the 2020 MMT Binospec observation (bottom panel, gray).
    The bottom panel shows the residuals of each spectrum and the ML prediction. }
    \label{f:spectra:comparison}
\end{figure*}

Archival data is a scan of a 4.7$\arcsec$ diameter circular aperture spectrum from the Mount Lemmon 1.5~m (60 inch) telescope taken on 4/5 February 1981, shown in the top panel in Figure \ref{f:spectra:comparison}, and a 6.1$\arcsec$ circular aperture spectrum taken using the Intensified Image Dissector Scanner on the KPNO 2.1~m, shown in the second panel in Figure \ref{f:spectra:comparison}.
The archival spectra are normalized by the maximum value in the spectrum for ease of comparison to the ML spectrum. The 60 Inch spectrum closely resembles an SDSS spectrum in aperture and signal to noise but is limited in wavelength coverage. The KPNO spectrum has a similar wavelength coverage but much lower signal-to-noise and a wider aperture compared to SDSS spectra.

Recent VIRUS-P IFU observations of Rubin's galaxy seek to reveal star-formation and abundances (Young et al. \textit{in prep}). A blue spectrum ($<$ 5800~\AA) was extracted using a 4.16$\arcsec$ aperture to compare with the ML prediction, shown in Figure \ref{f:spectra:comparison}, third panel. 

A targeted $3\times300$~s optical spectrum was obtained at the MMT Observatory with the instrument Binospec \citep{Fabricant19} in November 2020 with a position angle = 35$^\circ$ using the 270 lines/mm grating with a central wavelength of 6500~\AA. The observations were taken at RA=3:53:02.4811, DEC=+35:35:22.103 with a 1\arcsec slit. Data reduction was completed by the Binospec automated pipeline, which is an open-source IDL software package distributed under the GPLv3 license \citep[][\url{https://bitbucket.org/chil_sai/binospec}]{Chilingarian19}. This standard reduction resulted in the spectrum in Figure \ref{f:spectra:comparison}, fourth panel.

Both the VIRUS-P and the MMT/Binospec spectra have been renormalized using the maximum value in each spectrum's bandpass (VIRUS-P $4000-5700$~\AA, MMT $4000-7000$~\AA). This leaves a gradual difference in continuum between normalized ML and either spectrum. Subsequently, we  modeled this continuum difference between the ML prediction and the observed spectrum as a third degree polynomial to remove bandwidth-wide effects. We subtracted this continuum difference to facilitate a direct comparison between spectra and their features in Figure \ref{f:spectra:comparison}.

\begin{figure}
    \centering
    \includegraphics[width=\textwidth]{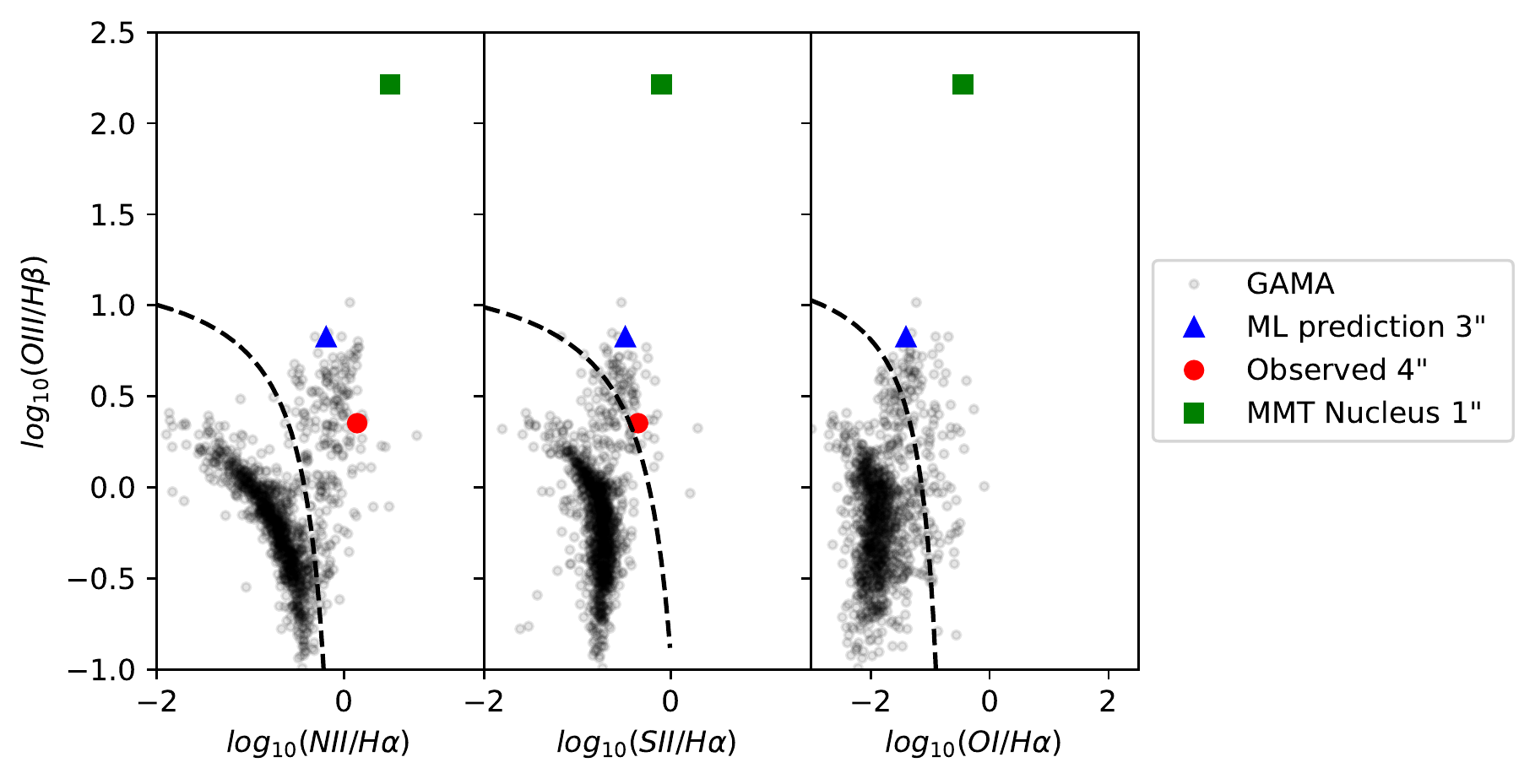}
    \caption{The BPT diagnostic diagram for nuclear activity. The Galaxy And Mass Assembly Survey data is shown for comparison with the \cite{Kauffmann03} star-formation/AGN demarcation (dashed lines). The observed line strengths reported in \cite{Keel83a} and the VIRUS-P observations for Rubin's galaxy put it in the active galaxies category. The ML predicted spectrum puts it in the AGN category. None of the line ratios prefer LINER or Seyfert classifications \citep{Kewley06}. The MMT line ratios are complicated as H$\beta$ is not well constrained, skewing the vertical position of the nucleus on the y-axis.}
    \label{f:BPT}
\end{figure}



\section{Discussion} \label{sec:discussion}

\subsection{How well did ML perform?}

Figure \ref{f:spectra:comparison} shows excellent agreement between the ML predicted spectrum and the observations.
Discrepancies in the stellar continuum are easily attributed to night sky lines (present in the 60 inch spectrum $>6800$\AA\ but not preserved in the SDSS vectorization). In the residual, one can identify a slight gradient in the KPNO spectrum's continuum ($<$5000~\AA), which is a similar calibration difference. 
Discrepancies in H$\alpha$/[{N}{ii}] and [{O}{iii}] line strengths are notable in Figure \ref{f:spectra:comparison}. 

The 60 inch and VIRUS-P spectra have closely comparable but not identical apertures to SDSS ($\sim4$\arcsec\ and 3\arcsec\ diameters respectively, Figure \ref{f:Hubble:nucleus}), and a slight offset is to be expected. Both archival spectra (Figure \ref{f:spectra:comparison}, top two panels) show excellent agreement with the ML predicted spectrum within their respective noise levels. The major differences are telluric features (e.g. $>6800$\AA\ in the 60 inch spectrum in Figure \ref{f:spectra:comparison}), not included in the original ML prediction (telluric features are a source of noise in the training set, and are therefore left out as much as practical).

The VIRUS-P spectrum broadly agrees well with the ML prediction. The only deviations are the [{O}{iii}] emission line and the {He}{i} absorption, which are both slightly under-predicted by the ML method (Figure \ref{f:spectra:comparison}, third panel). About a third of the equivalent width in the [{O}{iii}] emission and the {He}{i} absorption is missing (Figure \ref{f:spectra:comparison}, bottom panel). 

The MMT/Binospec spectrum (Figure \ref{f:spectra:comparison}, fourth panel) focuses on the very central region of Rubin's galaxy. The [{O}{iii}] and H$\alpha$/[{N}{ii}] emission line strengths, as well as the [{S}{ii}] absorption features at 4000~\AA, do appear to be under-predicted by the ML in comparison to the MMT spectrum. 


There are two factors resulting in differences between the ML predicted spectra and those observed: the aperture of the observed spectra and the aperture-distance effects in the SDSS training sample. 

Figure \ref{f:Hubble:nucleus} illustrates that the archival spectra and VIRUS-P are for a slightly larger aperture and the MMT one for a smaller aperture than an SDSS spectrum on this galaxy would have been. Stronger H$\beta$ and [{O}{iii}] from the nucleus can be explained by less dilution of the AGN signal. This can be seen as the progression of stronger observed emission lines in Figure \ref{f:spectra:comparison} with smaller apertures.

Secondly, the ML algorithm was trained on SDSS with a known aperture-distance effect (dilution of AGN signal in more distant galaxies), matching the spectra of more distant galaxies to Rubin's Galaxy's nucleus with a wider aperture diluting the AGN signal.

\subsection{AGN?}
\label{s:agn}

Identifying lower-power AGN can be difficult because selection effects are the most notable at the lower luminosities. Traditionally, one would select obscured or low-power AGN from IR colors rather than x-ray or emission lines in spectra. The ML approach offers an alternative, with Rubin's Galaxy an instructive example of one that could be identified using imaging alone.  

Figure \ref{f:BPT} shows where the observed and the ML predicted line strengths place the nucleus of Rubin's galaxy on the BPT diagram \citep{BPT}. 
The line strengths implied by the ML predicted spectrum imply strongly that Rubin's galaxy is an AGN (blue triangle in Figure \ref{f:BPT}). 

A combination of \cite{Keel83a} and VIRUS-P line strengths are more consistent with a lower-power AGN, closer to the dividing line between active galaxies and star-forming galaxies (red circle in Figure \ref{f:BPT}). The MMT/Binospec line strengths put the nucleus of Rubin's galaxy solidly in the AGN category (green square in Figure \ref{f:BPT}). The MMT spectrum does not constrain the H$\beta$ well and this skews the {O}{iii}/H$\beta$ ratio.
%
The observed or predicted line ratios are not consistent with those of either a LINER or Seyfert according to the \cite{Kewley06} criteria.

\begin{figure}
    \centering
    \includegraphics[width=\textwidth]{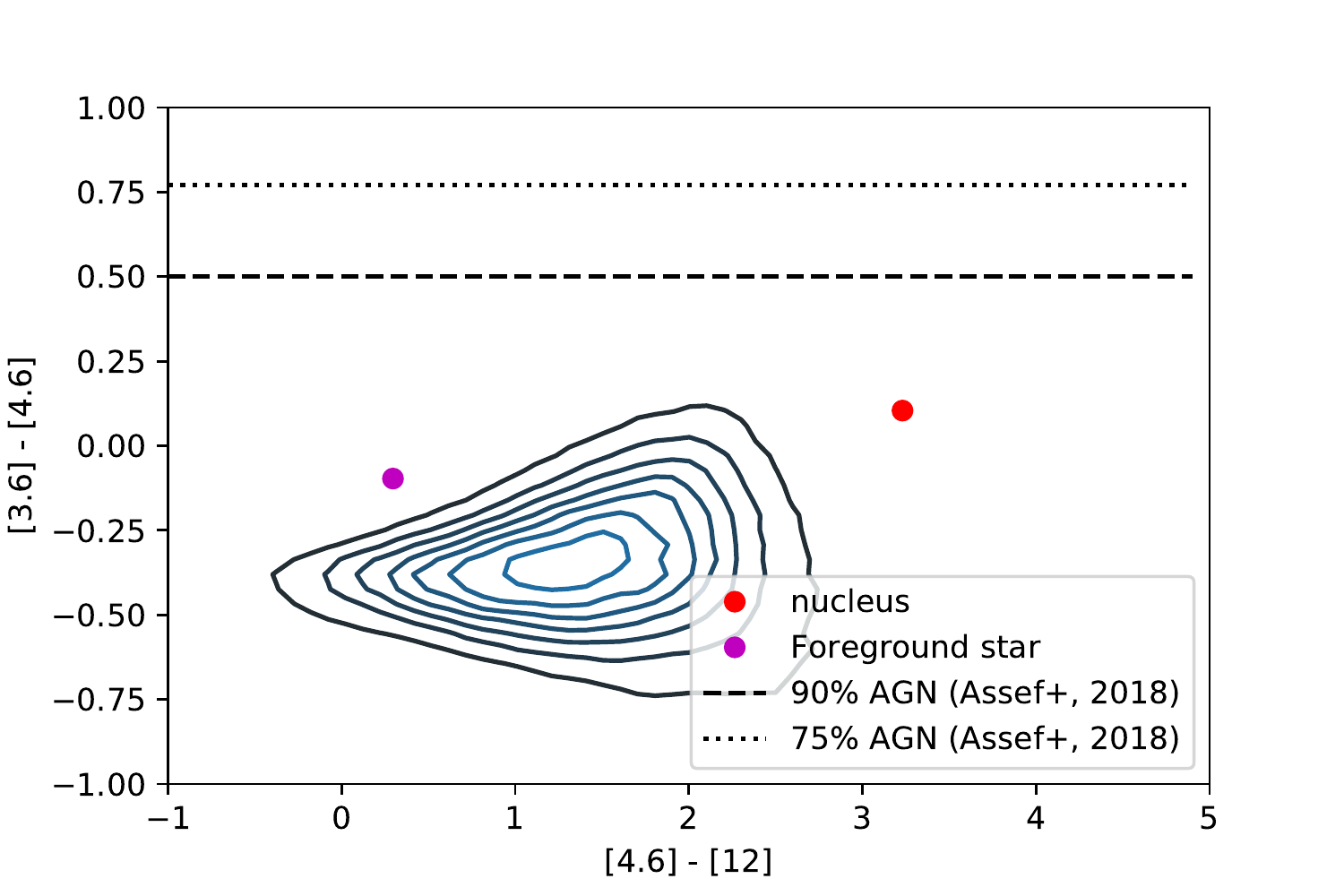}
    \caption{WISE colors of sources in the GAMA Equatorial Fields \citep{Cluver14}. The WISE colors of the ALLWISE catalog sources at the position of Rubin's Galaxy are shown. The colors of the bright foreground star projected on the disk of UGC 2885 are as would be expected. The colors of the nucleus are located away from the bulk of normal galaxies (contours) but the $[4.6]-[3.6]$ colour does not quite reach the AGN criterion of \cite{Assef18}. }
    \label{f:wise}
\end{figure}

Figure \ref{f:wise} shows the WISE colors of the central region of Rubin's galaxy for comparison with the  classifications of \citet{Jarrett11} and \citet{Cluver14}. 
The WISE colors are consistent with those of spiral, star-forming disk galaxies. The source catalog entry corresponding to the center of Rubin's Galaxy is in the part of color-space where star-forming galaxies border on obscured AGN and Seyfert galaxies, according to Figure 5 in \cite{Cluver14}. However the WISE $[3.6]-[4.5]$ color criterion from \cite{Assef15,Assef18} does not place Rubin's galaxy in the AGN bracket (dashed lines in Figure \ref{f:wise}). Given the angular resolution of WISE, dilution of the AGN signal by the PSF likely accounts for this lack of AGN identification i.e. the WISE flux is dominated by the stellar population of Rubin's galaxy. Based on the spectroscopic line strengths, the center of Rubin's galaxy is indeed an AGN.
WISE colors have the benefit of suffering less from dust obscuration allowing for unbiased searches of AGN in the nearby Universe. However, the WISE color section still misses AGN such as these and the ML approach may be a way to fill in the missing population.

\subsection{Discussion}

It is exciting to see that the ML prediction of an SDSS quality spectrum of this galaxy is so reliable, even when this galaxy resides in the space between star-forming and AGN. The ML method outperforms WISE color selection, the preferred method of identifying AGN in objects without a spectrum. A successful ML prediction in the loci of galaxy populations is reassuring but a tangible success for an extreme object is especially encouraging. How the ML algorithm transfers from e.g. Pan-STARRS broad-band \textit{grizy} images to the Rubin Observatory's LSST looks to be promising example of transfer learning in astronomy.


The combined line strengths of Rubin's Galaxy point to ongoing AGN activity. Combined with the very extended star-forming disk, the presence of an AGN makes this galaxy an interesting case: a massive disk galaxy, with no signs of interaction, relatively normal star-formation activity, and very regular rotation, yet using secular mechanisms only to fuel its central source.

\section{Conclusions}
\label{s:conclusions}


The ML predicted SDSS-quality spectrum from \cite{Wu20a} is remarkably close to the observed spectra (Figure \ref{f:spectra:comparison}). Deviations can be attributed to small differences in aperture or to residuals from telluric lines. 
The emission lines in the ML prediction are under-predicting the [{O}{iii}] and H$\alpha$ and [{N}{ii}] doublet compared to [{O}{ii}]. 
The successful identification of this galaxy's spectrum is a promising development in the use of ML. It is an encouraging development for future use on the Rubin Observatory data. 

In order to supply gas to the central black hole, the gas must lose angular momentum, and astronomers have long suspected that galaxy collisions (or at least interactions) provide the requisite torque to bring that gas into the nucleus \citep[e.g.][]{Hong15,Dietrich18, Gao20b, Marian20}. However, interactions are not the only avenue for fueling the AGN \citep[e.g.][]{McKernan10, Marian19}.  Rubin's galaxy is an example of an extremely isolated, undisturbed, and massive galaxy which shows clear evidence of nuclear activity, it may prove a useful laboratory to study the secular fueling processes on a grand scale. Further dissection of the inner regions of the galaxy may shed light on the secular mechanisms that supply fuel to AGN.

\section*{Acknowledgements}

The authors would like to thank the anonymous referee for their effort and time improving the manuscript. 
Observations reported here were obtained at the MMT Observatory, a joint facility of the University of Arizona and the Smithsonian Institution.
Support for this work was provided by NASA through grant number GO-15107 from the Space Telescope Science Institute, which is operated by AURA, Inc., under NASA contract NAS 5-26555.

\software{Astropy \citep{astropy}}

%

\clearpage

\end{document}